\documentclass[twocolumn,aps,prc]{revtex4}

\usepackage{graphicx}
\usepackage{dcolumn}
\usepackage{bm}
\newcommand{\be}{\ensuremath{\beta}}
\newcommand{\al}{\ensuremath{\alpha}}
\newcommand{\bp}{\ensuremath{\beta^+}}

\newcommand{\su}[1]{\ensuremath{^{#1}}}
\newcommand{\susub}[2]{\ensuremath{^{#1}_{#2}}}
\newcommand{\subemph}[2]{\ensuremath{#1}\ensuremath{_{#2}}}
\newcommand{\sub}[1]{\ensuremath{_{#1}}}
\newcommand{\g}{\ensuremath{\gamma}}
\newcommand{\et}{{\it et al.}}

\begin{document}


\title{\be-Decay Half-Life of the $rp$-Process Waiting Point Nuclide \su{84}Mo}

\author{J. B. Stoker$^{1,2}$,
P. F. Mantica$^{1,2}$,
D. Bazin$^{2}$,
A. Becerril$^{2,3,4}$,
J. S. Berryman$^{1,2}$,\\
H. L. Crawford$^{1,2}$,
A. Estrade$^{2,3,4}$,
C. J. Guess$^{2,3,4}$,
G. W. Hitt$^{2,3,4}$,
G. Lorusso$^{2,3,4}$,\\
M. Matos$^{2,4}$,
K. Minamisono$^{2}$,
F. Montes$^{2,4}$, 
J. Pereira$^{2}$,
G. Perdikakis$^{2}$,\\
H. Schatz$^{2,3,4}$,
K. Smith$^{2,3,4}$,
R. G. T. Zegers$^{2,3,4}$
}

\affiliation{$^{(1)}$
Department of Chemistry, Michigan State University,
East Lansing, Michigan 48824}
\affiliation{$^{(2)}$ 
National Superconducting Cyclotron
Laboratory, Michigan State University,
East Lansing, Michigan 48824} 
\affiliation{$^{(3)}$
Department of Physics and Astronomy, Michigan State University, 
East Lansing, Michigan 48824}
\affiliation{$^{(4)}$
Joint Institute for Nuclear Astrophysics, Michigan State University, 
East Lansing, Michigan 48824}

\date{\today}

\begin{abstract}
A half-life of 2.2 $\pm$ 0.2 s has been deduced for the ground-state 
\be{} decay of \su{84}Mo, more than 1$\sigma$
shorter than the previously adopted value.
\su{84}Mo is an even-even $N = Z$ nucleus lying on the proton dripline,
 created during explosive hydrogen burning in Type I X-ray bursts in the rapid proton capture ({\it rp}) process. 
The effect of the measured half-life on {\it rp}-process 
reaction flow is explored. Implications on theoretical treatments of 
nuclear deformation in \su{84}Mo are also discussed.

\end{abstract}

\maketitle

\section{\label{sec1:level1}Introduction}
\subsection{\label{sec1:level2}The Astrophysical $rp$-Process}
 The present model for the $rp$-process begins with the accretion of hydrogen and helium rich matter onto neutron stars from nearby companion stars. Gravitational energy is released in the form of X-rays as matter reaches the intimate regions of the neutron star. The matter is compressed as it forms an accretion disk and travels through the gravitational field gradient towards the neutron star, which eventually results in thermonuclear burning. The $rp$-process is a sequence of (p,\g) reactions and subsequent \bp{} decays up to \su{107,108}Te \cite{Schat04}. The process begins at \su{41}Sc, with the seed nuclei produced by a series of fast (\al,p) and (p,\g) reactions on CNO-cycle nuclei. The composition of the accreted matter governs the overall duration and peak temperature of the nuclear burning stage of X-ray bursts \cite{Schat01}.

The $rp$-process circumstances prove plausible for the synthesis of stable isotopes on the proton-rich side of the nuclear chart ($p$-nuclei) below $A = 107$. This is because the $rp$-process ashes are finally determined by \be{} decay from the proton-rich side of \be-stability \cite{Schat01}. The observed abundances for the $p$-nuclei \su{92,94}Mo and \su{96,98}Ru are currently at odds with abundance predictions using other solar production mechanisms \cite{Costa00}. The $rp$-process can produce these nuclei under certain circumstances, sustaining X-ray bursts as a possible contributor to their solar abundances. The amount of $rp$-process material emitted into the interstellar medium depends on the isotopic overproduction factor, the frequency of burst occurrences that produce $p$-nuclei, and the fraction of material that escapes the gravitational field of the neutron star \cite{Boyd02}. It has been pointed out that the high gravitational field in the vicinity of neutron stars hinders stellar ejection of all but the most energetic ash \cite{Weinb06}. The bulk of nuclei synthesized in the $rp$-process would not escape the gravitational field. This limited mass ejection is a major drawback for the modeling of X-ray bursts. Nevertheless, the amount of ejected material is still being debated and the composition of material produced in an $rp$-process event relies on experimental data, and in instances where these data are lacking, nuclear structure models \cite{Jose05}. The half-life of \su{84}Mo is a necessary experimental parameter for modeling reaction flow above mass 84, and directly determines the amount of \su{84}Sr formed during X-ray bursts.

\subsection{\label{sec1:level2b}Shape Deformation Along $N = Z$}
\begin{figure}
\centering
\includegraphics[width=0.5\textwidth]{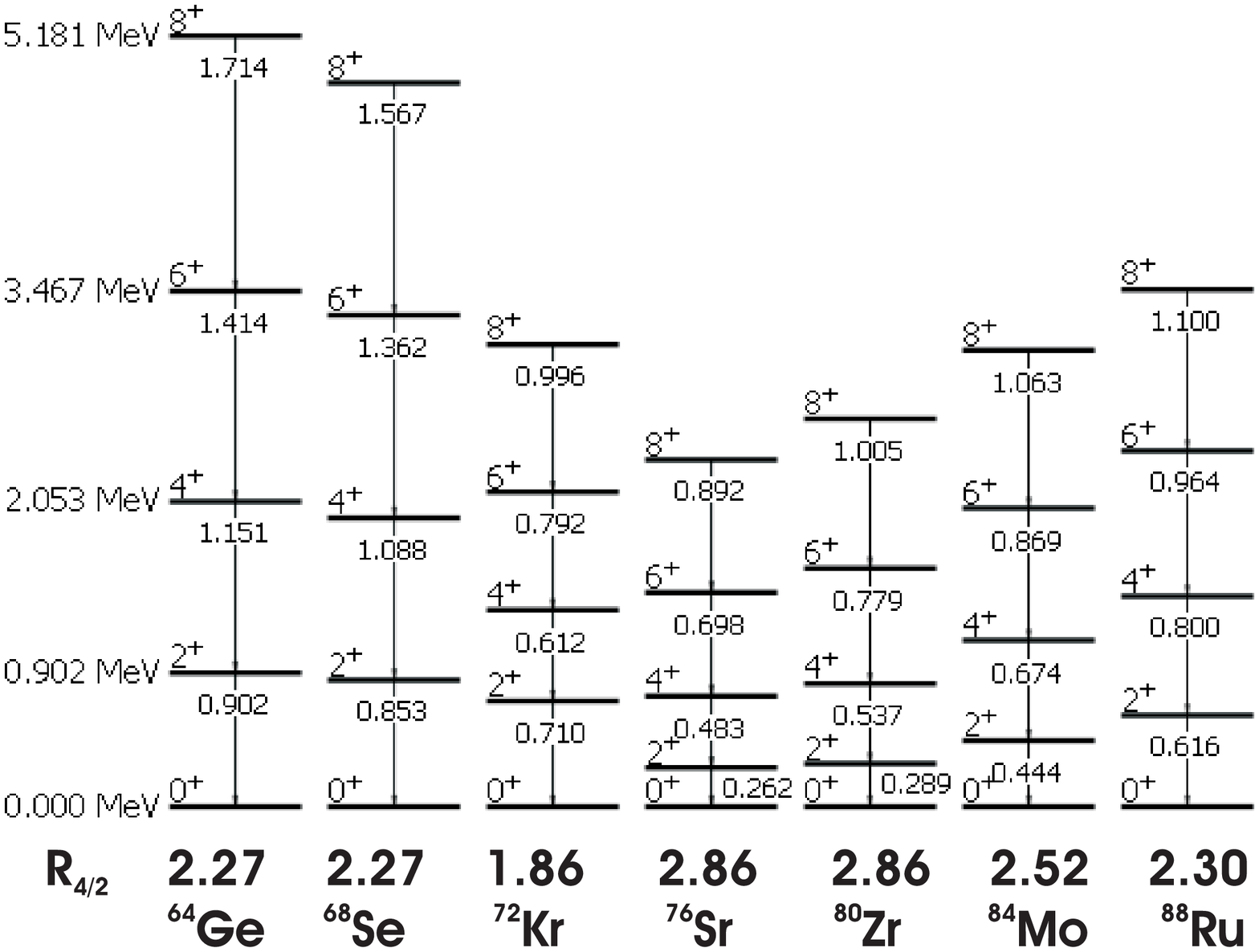}
\caption[Yrast States in Even-Even $N=Z$ Nuclei]{Yrast states up to 8\su{+} in even-even $N=Z$ nuclei. The transition energies are given in MeV. \subemph{R}{4/2} ratios are shown just below 0\su{+} state. Taken from \cite{Margi01,Margi04}.} 
\label{fig:N=Z_yrast}
\end{figure}

Deformed nuclei in general exhibit different single-particle level spacing and higher level densities than spherical nuclei. Single-particle levels, level densities and nuclear masses are important ingredients for calculating proton-capture rates and \be-decay half-lives \cite{Schat98}. The ratio of the $4\su{+}$ and $2\su{+}$ yrast state energies $[\subemph{R}{4/2} \equiv E(4\susub{+}{1})/E(2\susub{+}{1})]$ can be used as an indicator of shape deformation in even-even nuclei, with a smaller value representing a less deformed nucleus \cite{Caste93}. The \subemph{R}{4/2} trend for even-even nuclei along $N=Z$ reveals a deformation maximum at \su{76}Sr and \su{80}Zr (see Fig.\ \ref{fig:N=Z_yrast}) \cite{Bucur97,Margi01,Margi04}. A \subemph{R}{4/2} ratio of 2.52 for \su{84}Mo marks the beginning of a transition towards the believed spherical, doubly magic \su{100}Sn.

\begin{figure}[ht]
\centering
\includegraphics[width=0.5\textwidth]{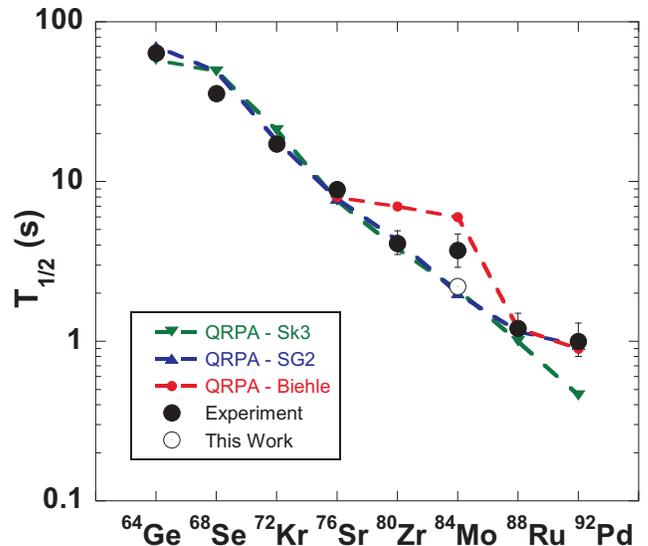}
\caption[QRPA Half-Lives Compared to Experiment]{(color online) Half-lives of even-even $N=Z$ nuclei for $A=64$ to $A=92$ (filled circles) compared to calculations using the QRPA.  Details of different theoretical self-consistent parameters are given in the text \cite{Biehl92, Sarri05}.  The newly measured result for \su{84}Mo is shown as the open circle.} 
\label{fig:half_lives}
\end{figure}

Theoretical predictions for the \be-decay half-life of \su{84}Mo within the Quasi-particle Random Phase Approximation (QRPA) vary from 2.0 s by Sarriguren \et{} \cite{Sarri05} to 6.0 s by Biehle \et{} \cite{Biehl92}. Fig.\ \ref{fig:half_lives} shows a comparison of the half-lives calculated by Sarriguren \et{} (QRPA-Sk3, QRPA-SG2) and Biehle \et{} (QRPA-Biehle) with experiment. The principle difference between these theoretical treatises is the set of nuclei used to calibrate the self-consistent interaction parameters for particle-particle coupling strength and nuclear deformation. 

The QRPA-Biehle prediction relies on the nearby nuclei \su{88,90}Mo, \su{92}Ru, and \su{94}Pd, which do not exhibit the same level (2.1 $\leq$  \subemph{R}{4/2} $\leq$ 2.23) \cite{Kapte76,Kabad94,Lingk97,Plett04} of deformation observed in the $N=Z$ region near $A=80$ but approach spherical shapes, lengthening the predicted half-lives. The self-consistent parameters for the QRPA-Sk3 and QRPA-SG2 cases were derived from experimental data from nuclei in the region of interest, and mostly reproduce the experimental half-lives using self-consistent deformations that minimize the energy. The previously measured half-life of \su{84}Mo reported by Kienle \et{} \cite{Kienl01} falls between the two calculations. This would imply a level of deformation unique to the mass region, perhaps inconsistent with the observed trend of measured $\subemph{R}{4/2}$ ratios along even-even $N = Z$ nuclei. 

Recent re-measurements of the half-lives of \su{80}Zr \cite{Ressl00} and \su{91,92,93}Rh \cite{Dean04} are shown in Table \ref{ta:half}. Each of these measurements are background suppressed by \g-ray gating and are systematically lower than those reported by Kienle \et{} \cite{Kienl01}, at times by more than 1$\sigma$. A similarly shorter result for a \su{84}Mo re-measurement would validate Sarriguren's approach for $N = Z$ nuclei in this region, and give consistency with the deformation implied by the measured $\subemph{R}{4/2}$. 

\begin{table}[ht]
\caption[Kienle \et Half-lives]{Experimental half-lives determined by Kienle \et{} \cite{Kienl01} compared with \be-\g{} correlated measurements \cite{Ressl00,Dean04}.}
\begin{ruledtabular}
\begin{tabular}{c c  c c c}
Nucleus & J\su{\pi} & T\sub{1/2} (s) \footnotemark[1] & & \be-\g coincidence T\sub{1/2} (s)\\
\hline \\
\su{80}Zr & (0\su{+}) & 5.3$\susub{+1.1}{-0.9}$ & & 4.1$\susub{+0.8}{-0.6}$ \footnotemark[2] \\
\su{93}Rh & (9/2\su{+}) & 13.9$\susub{+1.6}{-0.9}$ & & 11.9$\pm$0.7 \footnotemark[3] \\ 
\su{92}Rh & ($\geq$6\su{+}) & 5.6$\susub{+0.5}{-0.5}$ & & 4.66$\pm$0.25 \footnotemark[3]\\
\su{91}Rh & (9/2\su{+}) & 1.7$\susub{+0.2}{-0.2}$ & & 1.47$\pm$0.22 \footnotemark[3] \\
\end{tabular}
\footnotetext[1]{Ref.\ \protect\cite{Kienl01}.}
\footnotetext[2]{Ref.\ \protect\cite{Ressl00}.}
\footnotetext[3]{Ref.\ \protect\cite{Dean04}.}
\end{ruledtabular} 
\label{ta:half}
\end{table}

\section{\label{sec2:level1}Experimental Procedure}
Nuclei for this study were produced via fragmentation of \su{124}Xe projectiles accelerated to 140 MeV/nucleon in the coupled K500 and K1200 cyclotrons at National Superconducting Cyclotron Laboratory. The \su{124}Xe beam was impinged upon a 305 mg/cm\su{2} \su{9}Be target. Fragments were transported through the A1900 Fragment Separator for separation~\cite{Morri03}. The A1900 dipole settings were $B\rho$\sub{1,2} = 2.9493 Tm, and  $B\rho$\sub{3,4} = 2.5635 Tm with a 180 mg/cm\su{2} Al wedge placed at the intermediate image. Fragments within $\pm$ 0.5\% of the central momentum were transported to the A1900 focal plane.

\begin{figure}
\centering
\includegraphics[width = 0.5\textwidth]{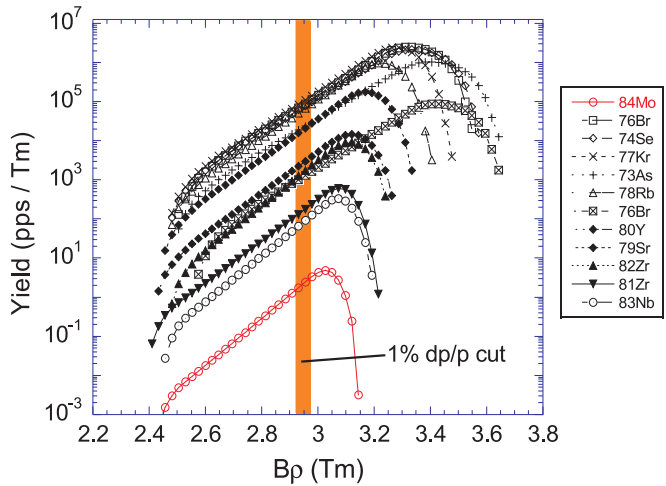}
\caption[LISE $B\rho$ Plot of Asymmetric Momentum Distribution]{(color online) Simulation of the yields in particles per second (pps) as a function of $B\rho$ for the principle isotopes created in this study.}
\label{fig:brhoAsymmetry07003}
\end{figure}

The momentum distributions of nuclei produced in intermediate-energy fragmentation reactions are asymmetric. In Fig.\ \ref{fig:brhoAsymmetry07003} is shown a simulation of yields as a function of $B\rho$ for the principle isotopes produced in this study. The simulation was performed with the program LISE \cite{Taras04}. The production rates from the low-momentum tails of more stable species exceeds that of the exotic \su{84}Mo such that a decay experiment would not be feasible without additional beam purification beyond that achieved with the A1900. 

\begin{figure}[htbp]
\centering
\includegraphics[width = 0.5\textwidth]{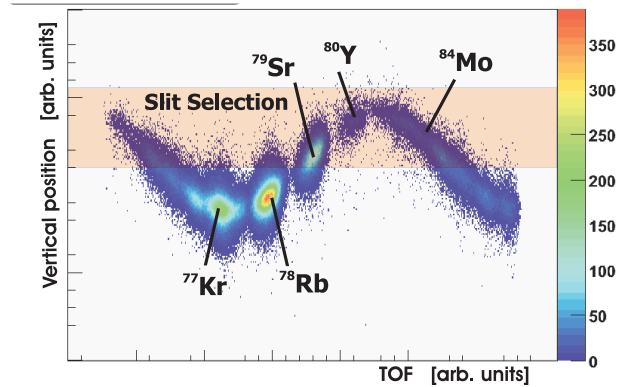}
\caption[RFFS Beam Sinusoid at Slit]{(color online) Plot of the vertical beam position as a function of TOF. The rectangle highlights the region accepted through the 10 mm slit setting.}
\label{fig:posYvRFTOF}
\end{figure}

The new Radio Frequency Fragment Separator (RFFS) was implemented at NSCL to purify beams of neutron-deficient nuclei~\cite{Gorel05}. The device consists of two horizontally parallel plates 1.5 m long, 10 cm wide, and 5 cm apart. A time-varying electric field is applied across the plates with an amplitude up to 100 kV peak-to-peak. A voltage of 47 kV peak-to-peak was applied for this study. This field varied sinusoidally at a frequency of 23.145 MH$z$, matched to the K1200 cyclotron. The time required for nuclei in the study to pass through the full 1.5-m length of the RFFS was roughly equal to 1/3 of a RF period. The Time-Of-Flight (TOF) separation of the beam species resulted in each fragment seeing a different 1/3 of the sinusoidal voltage cycle, and therefore receiving each a different vertical ``kick''. A diagnostic box was located 6 m downstream of the RFFS exit and contained an adjustable slit system and particle-counting detectors. Two retractable parallel-plane avalanche counters (PPACs) were placed along the beam path, one upstream of the slit and the other downstream. The vertical deflection of the fragments at the slit position is presented in Fig.\ \ref{fig:posYvRFTOF}. The deflection range selected by the 10 mm slit gap is also indicated. The phase of the RFFS can be delayed relative to the cyclotron RF such that the fragment of interest appears at a trough, peak, or node of the sinusoid as desired. 

\begin{table}[tbp]
\caption[Selective Rejection of RFFS]{Isotope specific and overall beam rejection rates. The rejection factor is reported as the ratio of RFFS\sub{off}/RFFS\sub{on} for individual fragment yields. Rejection factors larger than 1 indicate a reduced transmission with the RFFS on. Rates for specific isotopes are given in particles per second per particle nanoamperes of beam, normalized to the measured rate of \su{84}Mo. The beam intensity was attenuated when RFFS was off (V = 0 kV) to preserve Si detector longevity.}
\centering
\begin{ruledtabular}
\begin{tabular}{c c c c c}
              & \multicolumn{2}{c}{Rate\footnotemark[1]} &  & \\
\cline{2-3} 
              & V = 0 kV       & V = 47 kV      & Rejection & Half-life\\
              & 50~mm slit & 10~mm slit & Factor & \\
\hline \\
\su{84}Mo & 1 & 1 & 1   & $3.7^{+1.0}_{-0.8}$~s\\
\su{83}Nb & 15 & 16 & 1 & $4.1 \pm 0.3$~s\\
\su{82}Zr & 80 & 40 & 2 & $32 \pm 5$~s\\
\su{81}Zr & 20 & 10 & 2 & $5.3 \pm 0.5$~s\\
\su{80}Y & 130 & 200 & 0.6 & $30.1 \pm 0.5$~s\\
&&&& $4.8 \pm 0.3$~s\\
\su{79}Sr & 4000 & 85 & 47 & $2.25 \pm 0.10$~min \\
\su{78}Rb & 18700 & 0.4 & 46700 & $17.66 \pm 0.08$~min\\
&&&& $5.74 \pm 0.05$~min\\
\su{77}Kr & 13500 & 0.3 & 45000 & $77.4 \pm 0.6$~min\\
\su{76}Br & 1150 & 15 & 77 & $16.2 \pm 0.2$~h\\
&&&& $1.31 \pm 0.02$~s\\
\su{74}Se & 1980 & 5 & 400 & stable\\
\su{73}As & 700 & 630 & 1.1 & $80.30 \pm 0.06$~d\\ 
\hline
Sum\footnotemark[2] & 83 & 0.5 & 180 \\
$I_{beam}$            & 0.8 pnA & 10 pnA &  \\ 
\end{tabular}
\footnotetext[1] {Rates normalized to \su{84}Mo, 5 $\times$ 10\su{-4} pps/pnA}
\footnotetext[2] {Absolute rate in pps/pnA}
\end{ruledtabular}
\label{tab:RFFSreject}
\end{table}

Table~\ref{tab:RFFSreject} quantifies the overall and isotope specific rejection factors observed during this study. The table demonstrates the importance of distinguishing the device's selective rejection rate as opposed to the overall rejection rate. Selective rejection of key contaminants \su{78}Rb and \su{77}Kr was of order 10\su{4}, while the counting rate of the fragment of interest \su{84}Mo was not affected. The overall rejection factor was 180. Fragments that passed through the RFFS slit system were delivered to the experimental end station for analysis.

The Beta Counting System (BCS) \cite{Prisc03}  was used for event-by-event correlation of fragment implantations with their subsequent \be-decays. The system employed a Double-Sided Silicon Strip Detector (DSSD), a single 995 $\mu$m $\times$ 4 cm $\times$ 4 cm silicon wafer segmented into 40 1-mm strips in both the \emph{x} and \emph{y} dimension, providing 1600 individual pixels. The high degree of segmentation and beam filtering were both necessary to realize an average time between implantations into a single pixel that exceeded the correlation time, which is set in software based on the expected half-life of the isotope of interest.

\begin{figure}
\centering
\includegraphics [width = 0.5\textwidth]{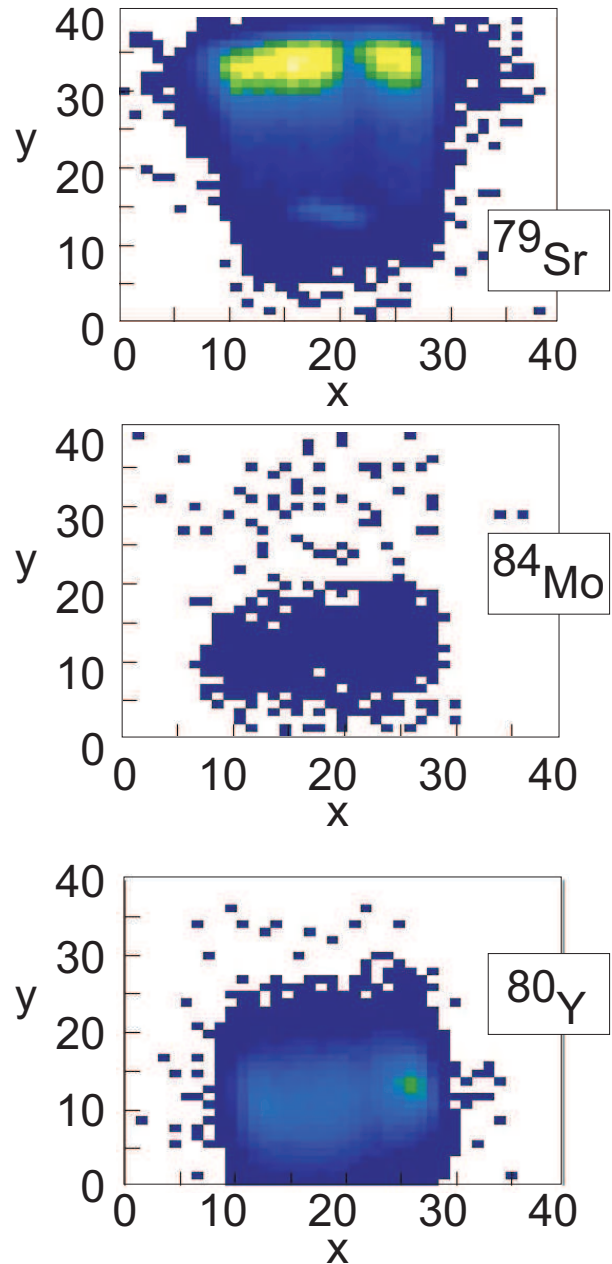}
\caption[DSSD distributions]{(color online) Distribution of \su{79}Sr, \su{84}Mo, and \su{80}Y fragments in the DSSD. The vertical deflection of fragments by the RFFS based on time-of-flight produces a non-uniform distribution of fragments in the $y$ dimension.} 
\label{fig:DSSDxy}
\end{figure}

The distribution of \su{84}Mo fragments on the surface of the DSSD is presented in Fig.\ \ref{fig:DSSDxy}.  As noted above, background is controlled in the experiment by taking advantage of the high degree of segmentation of the DSSD.  Here the \su{84}Mo fragments are distributed over $\sim 1/4$ of the DSSD surface, and overlap in position most strongly with \su{80}Y ions but not \su{79}Sr ions, where \su{80}Y and \su{79}Sr are two of the three significant contaminants in the secondary beam.  The other sizeable contaminant, \su{73}As, will have minimal impact on the accurate correlation of \su{84}Mo $\beta$-decay events, since its half-life is of order days (see Table \ref{tab:RFFSreject}).  The percentage of \su{84}Mo implantations that were preceeded by a \su{80}Y implantation in the same pixel was evaluated for the 10-s correlation time used to establish the decay curve for regression analysis as outlined below.   These so-called ``back-to-back'' implantations occurred for $\sim 8$\% of the \su{84}Mo implantation events, and were essentially distributed equally in time as the average time between implantations in a given pixel well exceeded the correlation time.  The impact of the ``back-to-back'' \su{84}Mo-\su{80}Y implantations is further mitigated by the fact that \su{80}Y has two $\beta$-decaying states as listed in Table \ref{tab:RFFSreject}.  The isomeric $1^-$ state of \su{80}Y decays with a half-life of $4.8 \pm 0.3$~s, and is a more significant contributor to the background analysis of the \su{84}Mo $\beta$ decay than the longer-lived $30.1 \pm 0.5$~s ground state with $J^\pi = 4^-$ \cite{Dorin99}.  $\beta$ decay of the  $1^-$ isomeric state contributed only $\sim$10\% to the total \su{80}Y activity, determined from analysis of the delayed $\gamma$-ray spectrum of \su{80}Y collected in the present work.     

Prompt and delayed $\gamma$ rays were monitored at the experimental end station with 16 Ge detectors from the Segmented Germanium Array (SeGA) \cite{Muell01}.  The detectors were mounted on a frame designed to closely pack the cylindrical Ge crystals in two rings of 8 detectors around the BCS chamber for \g-ray detection. Each detector was mounted with its cylindrical axis parallel to the beam axis. The DSSD was located in the plane that separated the upstream and downstream Ge detector rings, thereby maximizing the overall detection efficiency of \g-rays emitted from nuclei implanted in the DSSD.  

Ions implanted in the DSSD were monitored by location for positron emission in the implantation pixel, as well as the four nearest-neighbor pixels, over a correlation time of 10 s. A correlation efficiency for \be{} detection of $\approx$ 38\% was realized.

Three Si PIN detectors (PIN1-3) placed upstream of the DSSD served as active degraders. PIN1 was used for the $\Delta$E and TOF start signal during particle identification (PID). The active degrader thicknesses were selected such that fragments were stopped in the front 1/3 of the DSSD, increasing the probability of detecting the small $\Delta$E of a \be-particle emitted in the downstream direction. 

Downstream of the DSSD were six 5 cm $\times$ 5 cm Single-Sided Silicon Strip Detectors (SSSD1-6), which were primarily used to veto signals from light particle that were transported to the end station. Each SSSD was segmented into sixteen strips on one face and aligned so that the segmentation in each successive detector alternated in \emph{x} and \emph{y}. 

The DSSD provided a trigger for the master gate  on either an implantation or decay event that registered a coincidence signal in the front and back strips. SeGA readout was opened for 20 $\mu$s following each event trigger, enabling coincidence measurements of fragment-\g{} and \be-\g{} events. A coincidence register monitored signals from the other Si detectors in the BCS telescope, reducing dead time by signaling readout only for energy signals within the master gate.

\section{\label{sec3:level1}Results and Discussion}

\begin{figure}
\centering
\includegraphics[width = 0.5\textwidth]{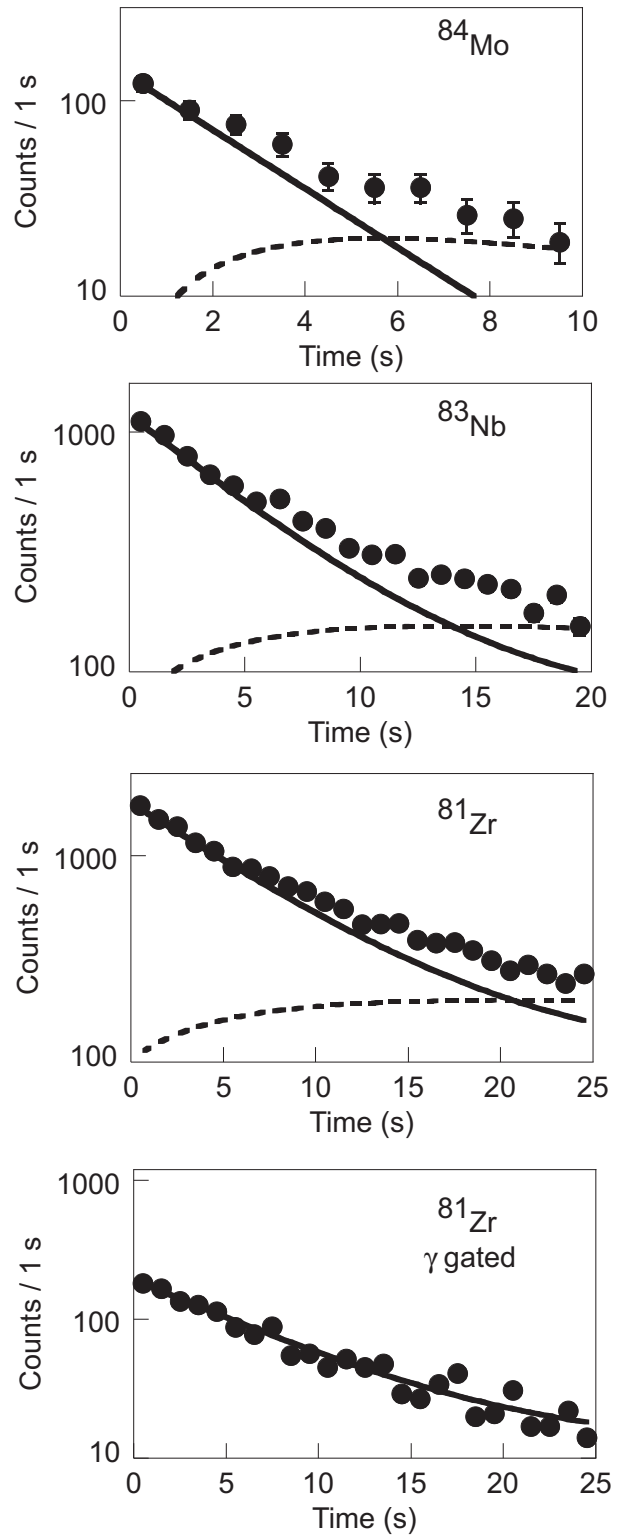}
\caption[84Mo Half Life Fit]{Decay curves for \su{84}Mo, \su{83}Nb, and \su{81}Zr.  
The solid and dashed lines are the parent and daughter contributions, respectively, 
to the decay and include a linear background.  Details of the fitting procedure are 
given in the text.}
\label{fig:84Mo}
\end{figure}

The decay curve for \be-decay events that occurred within 10 s in the same pixel or 4 nearest-neighbor pixels of a \su{84}Mo implantation is shown in Fig.\ \ref{fig:84Mo}. These data were fit based on the maximization of a Poisson probability log-likelihood  function that considered the exponential decay of the \su{84}Mo parent, the exponential growth and decay of the \su{84}Nb daughter using a T\sub{1/2} of 9.5 s \cite{Dorin99}, and a linear background.  The deduced half-life was 2.0 $\pm$ 0.4 s based on a sample size representing 1037 implantations. The reliability of the analysis approach is demonstrated by the reproduction of the known half-lives of \su{83}Nb and \su{81}Zr.  The half-life of \su{83}Nb was deduced by Kuroyanagi {\it et al.} \cite{Kuroy88} as $4.1 \pm 0.3$~s.  The value deduced from the decay curve in Fig.\ \ref{fig:84Mo} was $3.8 \pm 0.2$~s, consistent with the previous result.  A  half-life of $5.3 \pm 0.5$~s for \su{81}Zr was extracted by Huang {\it et al.} \cite{Huang97,Huang99} from a fit of a selective $\gamma p$-gated decay curve.  The fit of the decay curve for $\beta$ events correlated with \su{81}Zr implantations shown in Fig.\ \ref{fig:84Mo} resulted in a half-life value of $5.0 \pm 0.2$~s.  A decay curve for \su{81}Zr was also generated by gating on known $\gamma$ rays in the daughter \su{81}Y with energies 113, 175, and 230 keV, that were observed following $\beta$ decay for the first time.  A fit to this more selective decay curve, taking into account the exponential decay of the \su{81}Zr parent and a constant background, culminated in a half-life of $4.8 \pm 0.5$~s, again consistent with the the previously reported half-life and that deduced here considering only $\beta$-decay events correlated with \su{81}Zr implantations. 

One shortcoming of the fits performed with the Poisson probability log-likelihood function is that all decay times were considered to be independent, which does not accurately describe the probability density of daughter decays.  Therefore, decay events were also evaluated using a Maximum Likelihood analysis (MLH) \cite{Schne96} that considered \be-decay chains of up to 3 members.  Analysis of the \su{84}Mo decay chain considered events that occurred within 20 s in the same pixel of a \su{84}Mo implantation. Fixed values of 9.5 s and 25.9 min were used for the decay time parameters of the daughter and granddaughter, respectively. Background rates specific to each individual decay chain were determined by measuring the average rate of uncorrelated \be{} events on a pixel-by-pixel basis. A half-life of 2.2 $\pm$ 0.2 s was deduced using all 366 extracted chains. This value is consistent with the above described Poisson distribution maximum likelihood method, though the approach is inherently better because it relies on a probability density function that considers the dependencies of the parent, daughter, and granddaughter decays, as well as background. 

The new half-life value is more than 1$\sigma$ below the value of 3.7 \susub{+1.0}{-0.8} deduced from 37 implantations by Kienle \et{} \cite{Kienl01}. The new, shorter value continues the systematic trend observed in Table \ref{ta:half} relative to the previously adopted value. It is also consistent with the above discussed deformation assumed in the Sarriguren \et{} \cite{Sarri05} self-consistent treatment and affirms the trend towards a spherical \su{100}Sn implied by the measured \su{84}Mo $\subemph{R}{4/2}$ \cite{Margi01}.

\begin{figure}[htbp]
\centering
\includegraphics[width = 0.6\textwidth]{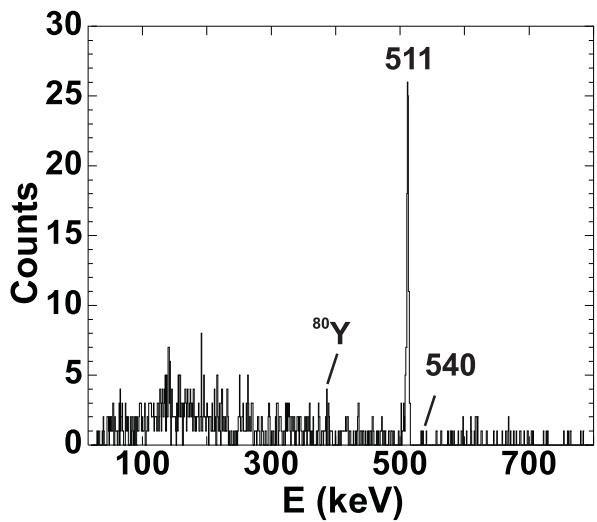}
\caption[84Mo Beta-gated gamma spec]{Spectrum of \g{} rays in coincidence with \be-decay events occurring in the same and the 4 nearest-neighbor pixels within 10 s of a \su{84}Mo implantation. The transition with energy 385.9 keV represents the decay of the 2\susub{+}{1} state in \su{80}Sr fed from \su{80}Y \be{}-decay, the principle source of background decay events.}
\label{fig:84Mo-Dgamma}
\end{figure}

The initial intent of the half-life remeasurement of \su{84}Mo was to gain additional selectivity in the analysis of the decay curve by $\gamma$-ray gating.  The energy spectrum for $\gamma$ rays found in coincidence with a \su{84}Mo \be{} decay is shown in Fig.\ \ref{fig:84Mo-Dgamma}. Unfortunately, no $\gamma$  rays were observed that could be attributed to the \su{84}Nb daughter.  Previous work has restricted the ground-state spin assignment for \su{84}Nb to 1, 2, or 3 with positive parity. A 1\su{+} \su{84}Nb ground-state would be consistent with the low \g{} intensity observed during \be{} decay from the \su{84}Mo 0\su{+} ground-state. Additionally, a 1\su{+} assignment would allow direct transitions into the the 0\su{+} \su{84}Zr ground state and explain the lack of evidence in Fig.\ \ref{fig:84Mo-Dgamma} for previously observed states populated via \su{84}Nb \be{} decay \cite{Dorin03}. Based on the simplistic assumption that 100\% of the \su{84}Nb \be-decays feed excited states in \su{84}Zr that depopulate through the known 2\susub{+}{1} level at 540.0 keV, we expected to observe a total of 14(9) counts at this energy in Fig.\ \ref{fig:84Mo-Dgamma}.  The absence of this 540.0 keV transition is consistent with a portion of the \su{84}Nb \be{} decay populating the \su{84}Zr ground state. This further supports a 1\su{+} spin and parity for the \su{84}Nb ground state.


\begin{figure}[htbp]
\centering
\includegraphics[width = 0.5\textwidth]{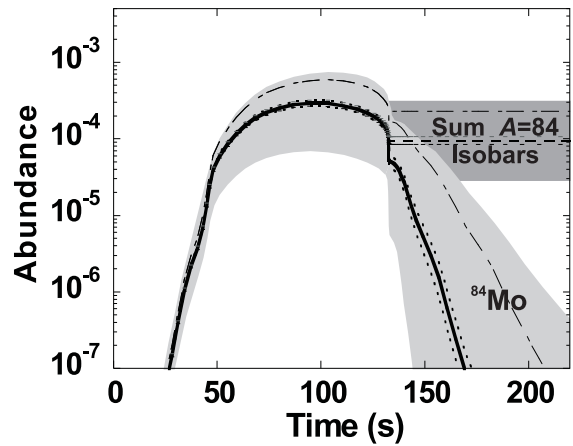}
\caption[Mass 84 Abundances]{Impact of \su{84}Mo half-life on the final $A = 84$ isobar abundance using a single zone X-ray burst calculation. Abundance is reported as a ratio of (mass fraction)/(mass number). Time 0 corresponds to the hydrogen-ignition start time. The solid line, bounded above and below with dotted uncertainties, shows the result using the newly-measured \su{84}Mo half life (T\sub{1/2} = 2.2$\pm$0.2 s). The dashed line corresponds to summed abundance of all $A = 84$ isobars. The dot-dashed line represents the yield calculated using the experimental upper limit of 4.7 s taken from the previously adopted \su{84}Mo T\sub{1/2}. Shaded regions highlight the range in abundance based on half-lives predicted previously (0.8 s$\leq$T\sub{1/2}$\leq$6.0 s).} 
\label{fig:rpCalc}
\end{figure}

The implications of the newly measured \su{84}Mo half-life on the $rp$-process were calculated using a single zone X-ray burst model based on ReaclibV1 rates provided by JINA Reaclib online database \cite{Reacl08}. The abundance with respect to burst duration is shown in Fig.\ \ref{fig:rpCalc} for \su{84}Mo (solid lines) and for all $A = 84$ isobars (dashed line). The shaded region covers the range of previously predicted half-lives (0.8 s $\leq$ T\sub{1/2} $\leq$ 6.0 s) given by various models \cite{Takah83,Biehl92,Sarri05}. The dot-dashed line represents the yield calculated using the experimental upper limit of 4.7 s taken from the previously adopted \su{84}Mo T\sub{1/2}. The order of magnitude uncertainty in the final \su{84}Sr abundance is reduced to less than a factor 2 with the new half-life. 

The amount of material processed above \su{56}Ni can become significant during X-ray bursts that last longer than 10 s. In such a scenario, the T\sub{1/2} of \su{84}Mo can have a considerable impact on the overall mass processing. A previous burst simulation \cite{Schat98} based on a low \al-separation energy for \su{84}Mo provided by the 1992 Finite Range Droplet Model predicted a Zr-Nb cycle via the reaction sequence
\begin{eqnarray}
\su{83}Nb(p,\al)\su{80}Zr(\bp)\su{80}Y(p,\g)\su{81}Zr \qquad \qquad \qquad \qquad \quad \nonumber \\
\qquad \qquad \qquad \qquad \bigg\{ 
\begin{array}{cc}
\!\!\!\!
(\bp)\su{81}Y(p,\g)\\
\!\!
(p,\g)\su{82}Nb(\bp)
\end{array}
\!\!
\bigg\} \
\su{82}Zr(p,\g)\su{83}Nb. 
\label{eq:ZrNbCycle} \nonumber
\end{eqnarray}
Escape from this cycle is only possible through \su{84}Mo \be-decay. The newly reported half-life is still longer than the 1.1 s value used for simulations in \cite{Schat98}, leading to a more pronounced bottleneck than predicted. The establishment of a Zr-Nb cycle would make the impact of the \su{84}Mo T\sub{1/2} for some $rp$-process scenarios considerable indeed. Measuring the \al-separation energy of \su{84}Mo is critical to determining the existence of such a cycle, and the real impact that it has on mass processing above $A = 84$.


\section{\label{sec4:level1}Conclusion}
The half-life for the $rp$-process waiting point nucleus \su{84}Mo has been re-measured to be 2.2 $\pm$ 0.2 s, more than 1$\sigma$ shorter than the previously adopted value. This new value is in line with the theoretical predictions of Sarriguren \et{} of the mid-mass $N = Z$ region consistent with a \su{84}Mo nucleus that begins a shape transition towards a spherical \su{100}Sn. \su{84}Mo is an important waiting point in the $rp$-process, determining mass abundance at and procession above $A = 84$. A measurement of the \su{84}Mo \al-separation energy is critical to determine the full impact of \su{84}Mo on the $rp$-process mass flow. \g-ray data were not sufficient to restrict the \su{84}Nb ground-state spin beyond what was previously reported, though a tentative 1\su{+} assignment is favored.

The experiment discussed in this paper was the first to use the NSCL RFFS. The TOF-specific beam rejection capability, critical for selectively reducing the overall beam rate, is shown to be sufficient to enable fragment-\be{} correlation of neutron-deficient, longer-lived species.

\begin{acknowledgments}
The authors thank the NSCL operations staff for providing the primary and secondary beams for this experiment and NSCL \g{} group for assistance setting up the Ge detectors from SeGA. We are appreciative of insightful statistics discussions with G.F. Grinyer and J. Thorpe. This work was supported in part by the National Science Foundation Grant Nos. PHY-06-06007 and PHY-05-20930 and JINA grant PHY-02-16783.
\end{acknowledgments}

\bibliography{prc}

\end{document}